# The Revival of White Holes as Small Bangs

Alon Retter[1] & Shlomo Heller[2]

[1] 86a/6 Hamacabim St., P.O. Box 4264, Shoham, 60850, Israel, alonretter@gmail.com

[2] Zuqim, 86833, Israel, shlomoheller@yahoo.com

___

**Abstract**

Black holes are extremely dense and compact objects from which light cannot escape. There is an overall consensus that black holes exist and many astronomical objects are identified with black holes. White holes were understood as the exact time reversal of black holes, therefore they should continuously throw away material. It is accepted, however, that a persistent ejection of mass leads to gravitational pressure, the formation of a black hole and thus to the "death of while holes". So far, no astronomical source has been successfully tagged a white hole. The only known white hole is the Big Bang which was instantaneous rather than continuous or long-lasting. We thus suggest that the emergence of a white hole, which we name a 'Small Bang', is spontaneous – all the matter is ejected at a single pulse. Unlike black holes, white holes cannot be continuously observed rather their effect can only be detected around the event itself. Gamma ray bursts are the most energetic explosions in the universe. Long γ-ray bursts were connected with supernova eruptions. There is a new group of γ-ray bursts, which are relatively close to Earth, but surprisingly lack any supernova emission. We propose identifying these bursts with white holes. White holes seem like the best explanation of γ-ray bursts that appear in voids. We also predict the detection of rare gigantic γ-ray bursts with energies much higher than typically observed.

Keywords: white holes, black holes, gamma ray bursts.

___

## 1. Introduction

In stars, there is a fine balance between the gravitational force, which operates towards the stellar center, and the sum of gas and radiation pressure, which conversely pushes outwards. In old stars the nuclear burning at the stellar core ceases and a collapse due to gravitation occurs. In neutron stars the electrons are compressed into the nucleus, combine with the protons and form neutrons. The collapse is then stopped by the pressure of the degenerate neutrons. For more massive stars, the contraction continues with no barrier and a black hole is formed. In black holes light is trapped inside the event horizon, which explains the origin of this term. In the absence of other known forces that can cease the collapse, it is believed that in black holes the material is concentrated in a single point with an infinite density or alternatively, that any discussion of the physical properties inside the event horizon is irrelevant. The single hypothetical central point of black holes is aliased a `singularity'. It is also accepted that the universe emerged from a singularity in the Big Bang about 13.7 billion (1.37 x$10^{10}$) years ago (Hubble 1929; Bennett et al. 2003; Spergel et al. 2007).

There is now extensive indirect observational evidence of the existence of black holes. Black holes are commonly found in binary systems and the orbital motion of the observed companion is used to estimate the mass of the obscured black hole (e.g. Ritter & Kolb 1998). It is even believed that most, perhaps all galaxies have a gigantic black hole at their center (Rees & Volonteri 2007). Certain observed types of supernova outbursts were identified with the collapsing end process of massive stars, and the supernova phenomenon was related with the detection of high energy γ-ray bursts as well (Woosley & Bloom 2006; Hjorth & Bloom 2011). The theoretical development of black holes led to the hypothesis that white holes should exist. Converse to matter-attracting black holes, in white holes material is spilled out of the singularity (Novikov 1965; Ne'eman 1965). These objects were described as 'lagging cores', parts of the initial Big Bang, long delayed in their expansion.

White holes were understood as the time reversal of black holes, and therefore, it was believed that they should persistently throw away matter, and be detected much easier than the dark black holes. Initially, white holes were even proposed as an explanation of the brightest objects in the universe - quasars and active galactic nuclei. It was concluded, however, that in white holes that continuously eject matter, a blue sheet of accreted highly accelerated matter will be formed at the event horizon of the white hole due to its gravitational force. The gravitational interaction between the white hole and the accreted material, will lead to an exponential suppression of the white hole emission. This will be the death of the white hole and a black hole will be generated (Eardley 1974; Ori & Poisson 1994). Indeed, while there is strong evidence of the presence of black holes, so far not a single astronomical object has been convincingly associated with a white hole. It can be said that white holes are literally dead.

## 2. The revival of white holes

It was proposed that the universe itself is a white hole (Gibbs 1998; Berman 2007). We believe that if the birth of the universe, the Big Bang, was a gigantic white hole, a weaker white hole, which we alias a 'Small Bang' should behave alike. The universe is believed to be born in a sudden moment. Similarly, we suggest that the appearance of a white hole should be immediate. Before this event happens, the white hole has no definite coordinates in space-time, and thus no gravitational effect on the environment of its destined eruption location. In contrast to previous claims, no white hole would spill out material for a long interval of time. Instead, like the birth of the universe itself, this process should be sudden, unpredicted and a single occasion event. Note that the best way to avoid the expected death of white holes is to eject all matter at a single instance (Blau 1989). We claim that white holes cannot be continuously observed as black holes. They can only be detected soon after they show up. We argue that white holes are not real astronomical objects. They are only a short window, through which material / energy is spilled out once. When a large amount of mass is ejected, the gravitational force, which is being expressed through mass, may eventually lead to a collapse and to the formation of a black hole, as was formerly concluded, but white holes cannot be destroyed after they are born, because their appearance is instantaneous. We believe that white holes are like birth – the baby can die, but not the birth itself.

How can white holes be detected? They should be highly energetic, with a large range in intensities, very fast and they can show up anywhere in space – close to Earth or at large distances. White holes can be born during the early stages of the universe, later on during the universe evolution and even today. In addition, they may occur both in dense or in isolated regions. All these requirements are met in γ-ray bursts.

## 3. Explaining a new class of Gamma-ray bursts by white holes

Gamma ray bursts (GRBs) were first detected about 40 years ago (Klebesadel et al. 1973). They are the most energetic explosions in the universe since the Big Bang. Many dozen models have been offered for the formation of GRBs, including white holes (Narlikar et al. 1974; Narlikar, & Appa Rao 1975; Ramadurai 2001). Extensive observations by satellites led to their division into two broad classes: short GRBs, whose γ-ray emission lasts less than two seconds and long GRBs, whose γ-ray radiation can be observed up to about one thousand seconds (Kouveliotou et al. 1993). Many long GRBs have been associated with supernova outbursts, and it is now accepted that long GRBs occur during the fatal stages of massive stars, when they explode and appear as supernovae. Short GRBs are explained by the merger of two compact stars - a neutron star with a black hole or with a second neutron star (Piran 1992; Panaitescu 2006). Long GRBs are more distant than short GRBs and thus carry more energy.

Recent observations suggest that there is a third new class of GRBs. The prototype of this group is GRB 060614. Its γ-ray emission lasted about 102 seconds similar to long GRBs, yet it shared many observational features with short GRBs. Despite its small red-shift, z=0.13, which means a relatively small distance to the observers, no supernova emission was seen from the source down to limits lower than any known supernova of its kind, and hundred of times fainter than the archetypical SN 1998bw (Galama et al. 1998). It was thus proposed that GRB 060614 as well as its twin GRB 060505 belong to a new class of bursters that require a novel model of explosion (Gal-Yam et al. 2006; Della Valle et al. 2006; Jakobsson & Fynbo 2007; Chattopadhyay et al. 2007; Caito et al. 2009). A new model for these GRBs was suggested, in which some cases of mergers between a massive white dwarf and a neutron star produce long-duration GRBs with no accompanying supernova (King et al. 2007). It was alternatively proposed that GRB 060614 might be produced by the tidal disruption of a star by an intermediate-mass black hole (Lu et al. 2008).

We propose that the new group of GRBs and many other GRBs can be explained by the instantaneous birth of white holes and the ejection of a large amount of matter. The white hole blast (or the probable formation of the black hole, which also happens in supernovae and compact stellar mergers) can account for the observed γ-ray emission as well as for other observational features of GRBs. White hole events are not connected with a supernova outburst, so this model easily explains the lack of supernova features in the members of the new class of GRBs. White holes seem like a natural explanation of GRBs that appear in isolated areas or voids, while merger scenarios between any various combinations of stars should only occur in populated areas. We thus further argue that supernova-less GRBs at high red-shifts, in which the lack of supernova emission is explained by poor observability, could be white holes as well.

For GRB 060614 the γ-ray fluence was 2 x $10^5$ ergs cm$^{-2}$, which corresponds to an isotropic γ-ray energy release of ~1 x $10^{51}$ ergs in the 1 keV - 10 MeV range in the GRB rest frame. For a radiation efficiency of 0.1, the isotropic kinetic energy is then 1 x $10^{52}$ ergs (Gehrels et al. 2006). This yields a low white hole mass of ~0.005 $M_\odot$, so this GRB may represent a white hole that emitted material of the order of some fraction of the solar mass. If indeed white holes exist, like black holes they could be found in all scales – from masses lower than the solar mass scale and up to a typical galaxy mass or even up to a few portions of the universe mass. If galaxy size white holes can blast they should eject all the mass at a single instance as the Big Bang and low mass white holes. It is, however, noted that it is very likely that similar to the distribution of black holes as a function of mass, the frequency of white holes decreases sharply with mass and maybe also with time since the Big Bang. This suggestion together with the proposed short window of white holes appearance may explain why gigantic GRBs are not regularly observed in the universe. They are expected to be very rare in comparison to those with a mass of order of the sun or a fraction of it.

## 4. Summary and conclusions

1.  We propose that white holes indicate the spontaneous ejection of energy / matter similar to the Bing Bang. Unlike black holes they cannot be continuously observed, rather they can only be detected around the event itself.

2.  White holes can appear anywhere and anytime, and only after they are born, there can be an interaction of the ejected matter with universal material. They are instantaneous and do not extend in time. Thus, the alleged instability of white holes and their extremely fast death because of the exponential growth of accreted matter around them are not precise.

3.  The observed wealth of different types of Gamma ray burst sources may contain appearance events of white holes. These should not be associated with supernova eruptions, and may occur both in galaxies and in voids. The new group of GRBs that are long, close and lack any supernova emission, is naturally explained by white hole blasts. The presence of GRBs with energies much larger than presently observed is predicted, but these could be quite rare.

## Acknowledgements

The anonymous referee is acknowledged for many wise comments which improved the paper.